\documentclass[useAMS,usenatbib]{mn2e}
\usepackage{graphicx,amsmath}
\usepackage{amssymb}
\usepackage{natbib}
\usepackage{microtype}
\usepackage{color}
\title[Low-$z$ ionizing background ]{Photon Underproduction Crisis: Are QSOs sufficient to resolve it?}
\author[Khaire and Srianand]{Vikram Khaire\thanks{E-mail:
vikramk@iucaa.ernet.in} and Raghunathan Srianand\thanks{E-mail: anand@iucaa.ernet.in}\\
IUCAA, Post Bag 4, Pune, India - 411007}

\begin{document}

\date{}

\pagerange{\pageref{firstpage}--\pageref{lastpage}} \pubyear{2015}

\maketitle

\label{firstpage}

\begin{abstract}
We investigate the recent claim of `photon underproduction crisis' by Kollmeier et al. (2014) 
which suggests that the known sources of ultra-violet (UV) radiation may not be sufficient 
to generate the inferred H~{\sc i} photoionization rate ($\Gamma_{\rm HI}$) in the low 
redshift inter-galactic medium. Using the updated QSO emissivities from the recent studies 
and our cosmological radiative transfer code developed to estimate the UV background, we show that the
QSO contributions to $\Gamma_{\rm HI}$ is higher by a factor $\sim$2 as compared to the 
previous estimates. Using self-consistently computed combinations of star formation rate 
density and dust attenuation, we show that a typical UV escape fraction of 4\% from 
star forming galaxies should be sufficient to explain the inferred $\Gamma_{\rm HI}$ by
Kollmeier et al. (2014). Interestingly, we find that the contribution from QSOs alone
can explain the recently inferred $\Gamma_{\rm HI}$ by Shull et al. (2015) which used the same 
observational data but different simulation. Therefore, we conclude that the crisis is not 
as severe as it was perceived before and there seems no need to look for alternate 
explanations such as low luminosity hidden QSOs or decaying dark matter particles. 
\end{abstract}


\section{Introduction}
Recently, \citet[][hereafter K14]{Kollmeier14}, used a cosmological hydrodynamic
simulation together with the latest measurements of the H~{\sc i} column density distribution, 
$f(N_{\rm HI})$, by \citet{Danforth14} in the low-$z$ intergalactic medium (IGM) and
reported a H~{\sc i} photoionization rate ($\Gamma_{\rm HI}$) at $z=0$. 
This is 5 times higher than the one (refer to as $\Gamma_{\rm HI}^{\rm HM}$) obtained from
the theoretical estimates of cosmic ultraviolet background (UVB) by \citet[hereafter HM12]{HM12}. 
This apparent discrepancy has led to the claim of a `photon underproduction crisis' suggesting that 
the origin of more than 80\% of H~{\sc i} ionizing photons is unknown and perhaps generated from 
non-standard sources.

For a given sight-line in a cosmological simulation, the inferred $f(N_{\rm HI})$
depends on the assumed $\Gamma_{\rm HI}$, the distribution of gas temperature and 
the clumping factor of the region producing the Ly-$\alpha$ absorption. The latter 
two quantities depend not only on the assumed initial power spectrum but also on various 
feedback processes that inject energy and momentum into the IGM from star forming galaxies. 
Therefore, the $\Gamma_{\rm HI}$ estimates using the $f(N_{\rm HI})$ will depend on how 
realistic the various feedbacks used in the simulation are. K14 have used 
the smooth particle hydrodynamics code {\sc gadget 2.0} \citep{Springel05} 
that includes feedback from galaxies in the form of momentum driven winds \citep{Oppenheimer08}. However, 
\citet{Dave10} suggested that these feedbacks produce negligible effect on 
$f(N_{\rm HI})$ for $N_{\rm HI}<10^{14}$ cm$^{-2}$. 

Recently, \citet{Shull15} have independently estimated $\Gamma_{\rm HI}$, using the 
same observed data but simulated spectra obtained using the grid based Eulerian N-body 
hydrodynamics code {\sc enzo} \citep{Bryan14}. They found a smaller $\Gamma_{\rm HI}$ 
than K14 but it is still a factor 2 higher than  $\Gamma_{\rm HI}^{\rm HM}$. They attributed 
the decrease in the derived $\Gamma_{\rm HI}$ as compared to K14 to the differences in the 
implementations of feedback processes in the simulations used. While \citet{Shull15} 
reduced the apparent tension, it still requires an appreciable contribution to the UVB 
from galaxies when one uses the previously estimated QSO emissivity.

In this study, we revisit the UVB calculations at $z\sim 0$ using the numerical radiative 
transfer code developed by us \citep{Khaire13} in line with  \citet{FG09} and HM12. We update the
QSO and galaxy emissivity and show that the QSOs alone can provide the $\Gamma_{\rm HI}$ 
inferred by \citet{Shull15} and to get the $\Gamma_{\rm HI}$ inferred by K14, we need only 
4\% of the ionizing photons to escape from galaxies (and not 15\% as suggested by K14) . 
Throughout this paper we use a cosmology with $\Omega_{\Lambda}=0.7$, $\Omega_{m}=0.3$ 
and $H_{0}=70$ km s$^{-1}$ Mpc$^{-1}$. 

\section{The radiative transfer}\label{sec.uvb}
Following the standard procedure \citep{Miralda90, Shapiro94, HM96, Fardal98, Shull99}, 
the average specific intensity, $J_{\nu_0}$ (in units of erg cm$^{\text{-2}}$ s$^{\text{-1}}$ 
Hz$^{\text{-1}}$ sr$^{\text{-1}}$), of the UVB at a frequency $\nu_0$ and redshift $z_0$ is given by, 
%
\begin{equation}\label{Eq.uvb}
J_{\nu_{0}}(z_{0})=\frac{1}{4\pi}\int_{z_{0}}^{\infty}dz\,\frac{dl}{dz}\,
(1+z_{0})^{3}\,\epsilon_{\nu}(z)\,e^{-\tau_{\rm eff}(\nu_{0},z_{0},z)}.
\end{equation}
%
Here, $\frac{dl}{dz}$ is the Friedmann-Lema\^itre-Robertson-Walker line element, $\epsilon_{\nu}(z)$ is the
comoving specific emissivity of the sources and  $\tau_{\rm eff}$ is an average effective optical 
depth encountered by photons of frequency $\nu_0$ at a redshift $z_0$ which were emitted from 
a redshift $z\,>z_0$ with a frequency $\nu\,>\nu_0$. The frequency $\nu$ and $\nu_0$ are related by 
$\nu=\nu_0(1+z)/(1+z_0)$. Assuming that the IGM clouds of 
neutral hydrogen column density, $N_{\rm HI}$, are Poisson-distributed along the line of sight, 
$\tau_{\rm eff}$ can be written as \citep[see][]{Paresce},
%
\begin{equation}\label{Eq.taueff}
\tau_{\rm eff}(\nu_{0}, z_{0}, z)=\int_{z_{0}}^{z}dz'\int_{0}^{\infty}dN_{\rm HI}
f(N_{\rm HI}, z) (1-e^{-\tau_{\nu'}})\,.
\end{equation}
%
Here, $f(N_{\rm HI}, z)$ is the number of H~{\sc i} clouds per unit redshift and column density 
interval having column density $N_{\rm HI}$. The continuum optical depth $\tau_{\nu'}$ is given by
%
$\tau_{\nu'}=N_{\rm HI}{\sigma_{\rm HI}(\nu')}+N_{\rm HeI}{\sigma_{\rm HeI}(\nu')}
+N_{\rm HeII}{\sigma_{\rm HeII}(\nu')},$
%
where, $N_i$ and $\sigma_{i}$ are the column density and photoionization cross-section, 
respectively, for species $i$ and $\nu'=\nu_{0}(1+z')/(1+z_{0})$. 

We use the same $f(N_{\rm HI}, z)$ used by HM12 and neglect the contribution of He~{\sc i} 
to $\tau_{\nu}$ because of its negligible abundance at $z<6$. We calculate $\tau_{\rm eff}$ 
following the prescription given in HM12. In the following section we provide the updated
source emissivity.
%
\begin{figure*}
\centering
    \includegraphics[width=15cm,keepaspectratio,clip=true]{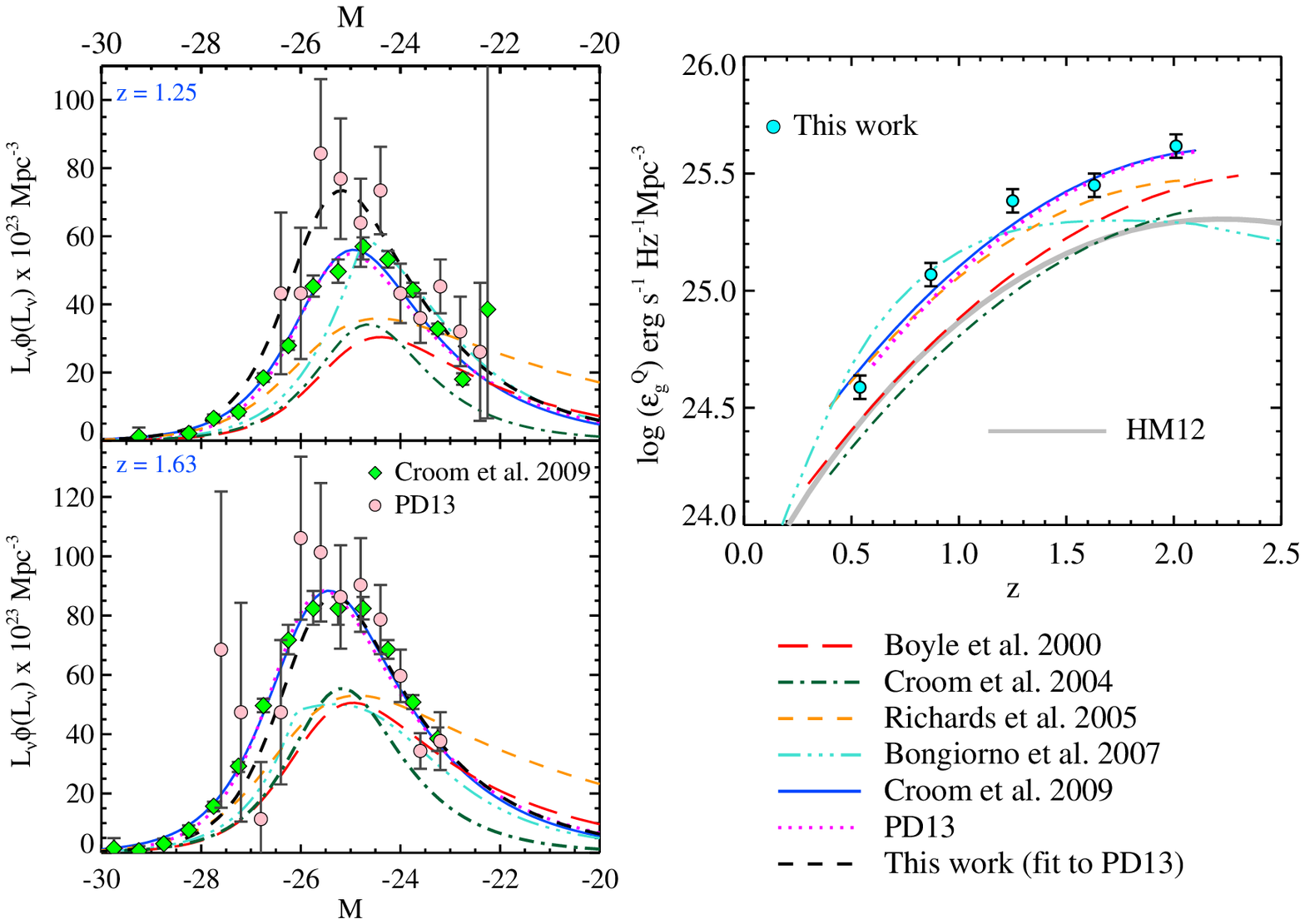}
\caption{ The $L_{\nu}\phi(L_{\nu})$ is plotted against M at g-band for two redshifts using 
various QLF reported in the literature. The area under each curve is proportional to the 
$\epsilon^Q_g$ (\emph{left panel}). The $\epsilon^Q_g$ obtained for these QLFs, our $\epsilon^Q_g$ from the 
fit to the PD13 and C09 QLF (cyan points; see table 1 for details) and the $\epsilon^Q_g(z)$ inferred 
from $\epsilon^Q_{912}(z)$ of HM12 are also plotted (\emph{right panel}). Here, the best fit PLE models 
of \citet{Boyle00}, \citet{Croom04}, with 2SLAQ data of \citet{Richards05}, C09 and PD13 
and the luminosity dependent density evolution model of \citet{Bongiorno07} are used. }
\label{fig.qlf}
\end{figure*}
%
\section{Emissivity of radiating sources}\label{sec.emis}
We calculate the UVB assuming only QSOs and galaxies are sources of the UV radiation.
Therefore, $\epsilon_{\nu}(z)=\epsilon^Q_{\nu}(z) + \epsilon_{\nu}^G(z)$ where 
$\epsilon^Q_{\nu}(z)$ and $\epsilon_{\nu}^G(z)$
are the comoving specific emissivity from QSOs and galaxies, respectively. 
\subsection{Comoving QSO emissivity}\label{sec.qso}
The $\epsilon^Q_{\nu}(z)$, in units of erg s$^{-1}$ Hz$^{-1}$ Mpc$^{-3}$, using the observed QSO 
luminosity function (QLF) at a frequency $\nu$ is given by
%
\begin{equation}\label{Eq.rho}
\epsilon^Q_{\nu}(z)=\int_{L_{\nu}^{min}}^{\infty} {L_{\nu}(z) \phi(L_{\nu}, z) dL_{\nu}}\,,
\end{equation}
%
where, $\phi(L_{\nu}, z)$ is the QLF at $z$ given in terms of specific luminosity
$L_{\nu}$ by
%
\begin{equation}\label{Eq.qlf1}
\phi(L_{\nu})= (\phi_L^*/L^*)\,[\,(L_{\nu}/L^*)^{-\gamma_1}+(L_{\nu}/L^*)^{-\gamma_2}\,]^{-1},
\end{equation}
%
and using the absolute AB magnitudes, M, by  
%
\begin{equation}\label{Eq.qlf2}
\phi(M)={\phi_M^{*}}[{10^{0.4(\gamma_1 + 1)(M-M^*)}+10^{0.4(\gamma_2 + 1)(M-M^*)}}]^{-1},
\end{equation}
%
where $\phi_M^{*}=0.921\phi_L^*$. We use $L_{\nu}^{min}=0.01L^*$ to calculate the $\epsilon^Q_{\nu}$. 
Note that, for most of the QLFs at $z<3.5$ the faint end slope  
$\gamma_1 \ge -1.4$ (see Table 2), for which the value of $\epsilon^Q_{\nu}$ is insensitive to the values of 
$L_{\nu}^{min}<0.01L^*$. For example, the maximum difference in the $\epsilon^Q_{\nu}$ calculated for 
$L_{\nu}^{min}=0$ and $0.01L^*$ is less than 5\% for $\gamma_1 \ge -1.4$ and $\gamma_2=-3.5$.

In Fig.1 (\emph{left panel}), we plot the $L_{\nu}\phi(L_{\nu})$ estimates against $g$-band magnitudes 
from various studies at two different $z$. The area under each curve is proportional to
the respective emissivity at $g$-band ($\epsilon^Q_{g}$).
It is clear from the Fig.1 that, as compared to old QLF measurements of 
\citet{Boyle00} and \citet{Croom04}, using the new measurements of \citet[][hereafter C09]{Croom09} and
\citet[][hereafter PD13]{Palanque13} will give a larger $\epsilon^Q_{g}$. This is indeed the case, as 
demonstrated in the \emph{right panel} of Fig.1 where we plot the $\epsilon^Q_{g}$ for these QLF measurements.
We have also plotted the $\epsilon^Q_{g}$ converted from the $\epsilon^Q_{912}(z)$ given at the H~{\sc i} Lyman limit 
(i.e at 912\AA) by HM12 using the relation $\log(\epsilon^Q_{g})= \log(\epsilon^Q_{912})+0.487$ which is
consistent with the spectral energy distribution (SED) used by HM12. 
This $\epsilon^Q_{g}$ is consistent with \citet{Boyle00} and \citet{Croom04} which is 
smaller by factor $\sim$ 1.5 to 2 as compared to C09 and PD13.

In our study, we use the latest QLF measurements as summarized in Table~\ref{lf_data}.
The first and second columns give the reference and the wavelength ($\lambda_{\rm band}$) at 
which the QLF is reported, respectively. For $0.3<z<3.5$, in each redshift bin we fit the observed QLF 
with the form given in Eq.~\ref{Eq.qlf2} using an {\sc idl mpfit} routine by fixing the values of $\gamma_1$ and 
$\gamma_2$ to those reported in the respective references (our fits are also presented 
in Fig~\ref{fig.qlf}). We use our best fit $\phi^*_M$ and $M^*$ to obtain $\epsilon_{\lambda{\rm band}}(z)$ 
(see Table~\ref{lf_data}). At other redshifts, we take the best fit QLF parameters given in the 
respective references and calculate the $\epsilon^Q_{\lambda{\rm band}}(z)$. 
In Fig 1 (\emph{right panel}), we show that the $\epsilon^Q_{g}(z)$ at $z<2$ obtained using our fit is 
consistent with the pure luminosity evolution (PLE) models of PD13 and C09.

We convert the $\epsilon^Q_{\lambda{\rm band}}(z)$ into $\epsilon^Q_{912}(z)$ 
using the broken power law QSO SED $\rm L_{\nu}\propto \nu^{-\alpha}$, which we adopt
for our UVB calculations. In the soft X-ray regime above energy 0.5 keV ($\lambda \le \rm 24.8\AA$) 
we use $\alpha=0.9$ \citep{Nandra94}. Following \citet{Stevans14}, we use $\alpha=1.4$ for 
$24.8 <\lambda \le \rm 1000\AA~$\footnote{\citet{Stevans14} found 
$\alpha=1.41\pm0.15$ using QSO composite spectrum down to about 500\rm \AA. We extrapolate it upto
$\lambda\sim25\rm \AA$. Note that the UVB at $\lambda< 500\rm \AA$ has negligible contribution 
to $\Gamma_{\rm HI}$.} and $\alpha=0.8$ for $1000 < \lambda \le \rm 2000\AA$. 
For $\lambda> \rm 2000\AA$ we use $\alpha=0.5$.

For the observed QLF at $z<3.5$, the SEDs used to perform continuum $K$-corrections
in the original references ($\rm L_{\nu}\propto \nu^{-\alpha'} $; $\alpha'$ is given in the last
column of Table~\ref{lf_data}) are different from our adopted SED at $\lambda\ge 1500$\AA.
For consistency, we recompute the specific emissivity $\epsilon^Q_{\lambda{\rm rest}}(z)$ at 
$\lambda_{\rm rest} = \lambda_{\rm band}/(1+z)$ using $\alpha'$ from the corresponding reference 
and then use our adopted SED to convert $\epsilon^Q_{\lambda{\rm rest}}(z)$ in to $\epsilon^Q_{912}(z)$. 
This is not needed for $z\ge 4$ where the QLFs are obtained at 1450\AA\ in the QSO's rest 
frame using appropriately matched filters without applying additional $K$-corrections. 
The errors on the $\epsilon^Q_{912}(z)$ given in the Table 1 are the maximum and minimum
difference we get using the errors in $\gamma_1$ and $\gamma_2$ given in original references
except at $z=0.15$. In this case the error on $\epsilon^Q_{912}$ is the difference we 
get in $\epsilon^Q_{912}$ if we use $L_{\nu}^{min}=0.1$ and 0.001. 

\begin{figure*}
\centering
    \includegraphics[bb=70 370 550 710, width=10cm,keepaspectratio,clip=true]{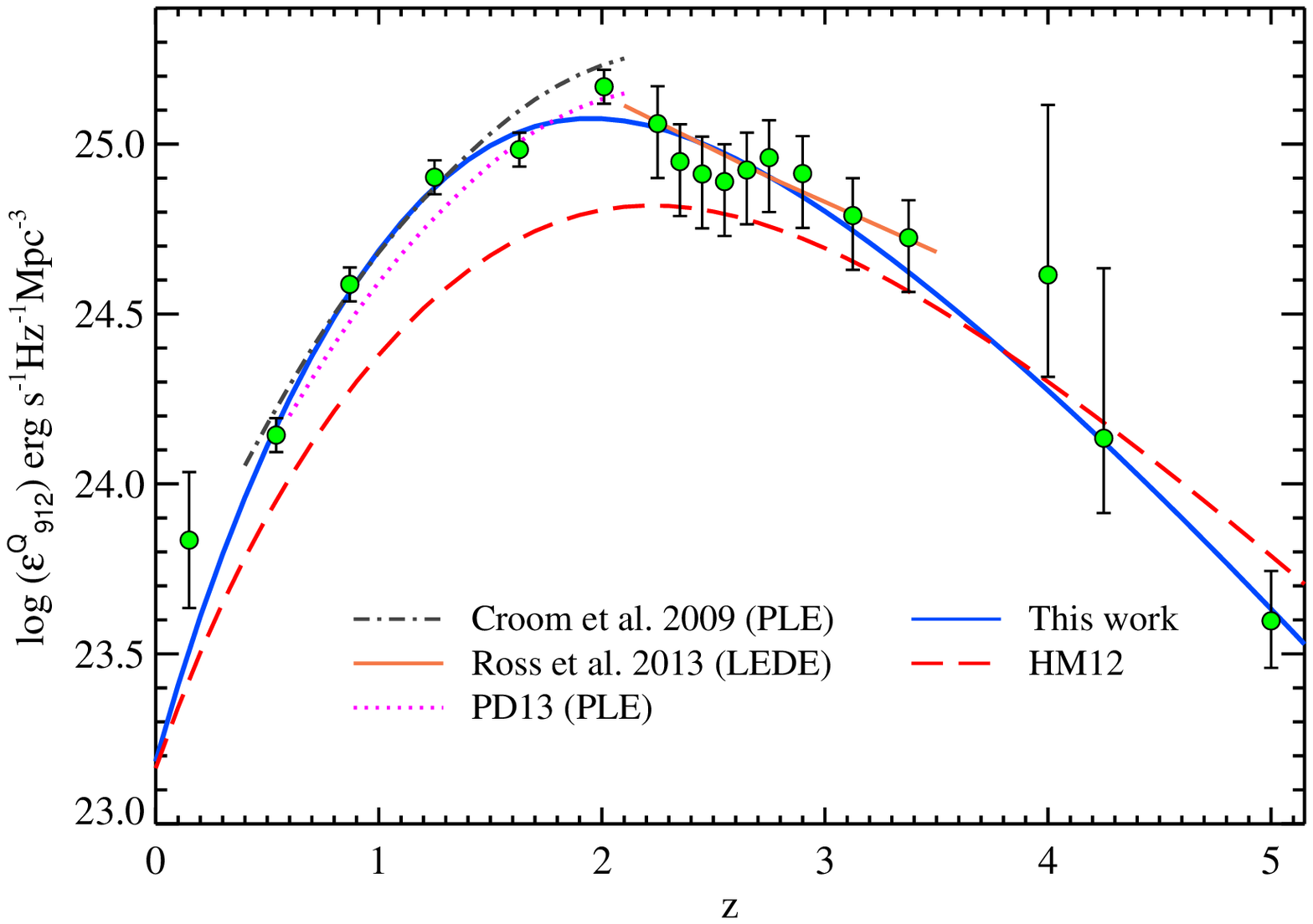}
\caption{The best fit comoving ${\epsilon^Q_{912}(z)}$ used in this paper (\emph{blue curve}) along with 
the ${\epsilon^Q_{912}(z)}$ used by HM12. The \emph{green circles} are the $\epsilon^Q_{912}$ given in Table 1.
The \emph{blue curve} is simply a fit to these points. The $\epsilon^Q_{912}(z)$ obtained using 
the PLE model of C09 (\emph{dot dash curve}) and PD13 (\emph{dotted curve}) and the LEDE model of 
\citet[][orange curve]{Ross13} is shown.} 
\label{fig.1}
\end{figure*}

All our $\epsilon^Q_{912}(z)$ measurements as a function of $z$ are plotted in Fig.~\ref{fig.1}. 
We fit these points using a functional form similar to that of HM12 and obtain the following best fit,   
%
\begin{equation}\label{Eq.Eqso}
\epsilon^Q_{912}(z)=10^{24.6}\,(1+z)^{5.9}\,\frac{\exp(-0.36z)}{\exp(2.2z) + 25.1}\,\,.
\end{equation}
%
For comparison, in Fig.\ref{fig.1}, we show this best fit ${\epsilon^Q_{912}(z)}$ 
along with the ${\epsilon^Q_{912}(z)}$ used by HM12. For $z<3.5$, our ${\epsilon^Q_{912}(z)}$ is higher than 
that of HM12 and the maximum difference of factor 2.1 occurs at $z\sim 1.5$. The peak in
${\epsilon^Q_{912}(z)}$ also changes from $z=2.2$ from HM12 to $z=1.95$.
In Fig.~\ref{fig.1}, we also show the $\epsilon^Q_{912}(z)$ obtained using the PLE models of C09 and PD13 and the 
luminosity evolution and density evolution (LEDE) model of \citet{Ross13}. These are consistent with 
our fit in Eq.\ref{Eq.Eqso}. Note that the PLE models of C09 and PD13 give identical values of $\epsilon^Q_{g}(z)$ 
(see \emph{right panel} of Fig.1) but differ slightly in the $\epsilon^Q_{912}(z)$ since they use 
different SED for continuum $K$-correction (see Table 1).
%
%
%
%
\subsection{Comoving galaxy emissivity}\label{sec.gal}
In \citet[][hereafter KS14]{Khaire14}, by matching the observed galaxy emissivity from 
multi-band, multi-epoch galaxy luminosity functions, we have determined self-consistent 
combinations of the star formation rate density (SFRD) and 
dust attenuation magnitude in the FUV band ($A_{\rm FUV}$)
for five well known extinction curves. It has been found that the SFRD($z$) and $A_{\rm FUV}(z)$ 
estimated using the average extinction curve of the Large Magellanic Cloud Supershell (LMC2) 
is consistent with various observations.

Here, as our fiducial model, we use the $\epsilon^G_{\nu}(z)$ computed from the SFRD($z$) 
and $A_{\rm FUV}(z)$ obtained in KS14 for the LMC2 extinction curve (see tables 2 and 4 in KS14).
Our SFRD at $z<0.5$ is a factor $\sim3$ higher than that of HM12 \citep[see also][]{Madau14}. 
However, the difference decreases at higher $z$, and becomes less than 10\% at $z\sim3$.
A small fraction, $f_{\rm esc}$, of the generated H~{\sc i}
ionizing photons ($\lambda <912$\AA) from the stellar population are assumed to 
escape through holes in galaxies 
(i.e by assuming that dust does not modify the SED at $\lambda<912$\AA).
We assume that there are no He~{\sc ii} ionizing photons ($\lambda \le 228$\AA) 
escaping the galaxy. This is a reasonable assumption in the $z$-range of our interest.  
We approximate the galaxy emissivity at $\lambda<912$\AA\ with a power-law 
$\epsilon^G_{\nu}\propto \nu^{-1.8}$. The exponent is fixed to reproduce the $\Gamma_{\rm HI}$ 
obtained from the model spectrum itself. Note that the exponent and the total H~{\sc i} ionizing 
photons generated inside the galaxy depends on the metallicity, initial mass function (IMF),
stellar rotation rates and adopted evolutionary tracks \citep[see][]{Topping15}.
In our galaxy models obtained from {\sc starburst99} \citep{Leitherer99}, we use the Salpeter IMF with 
$0.4$ times solar metallicity. See KS14 for a discussion on the
uncertainties in estimating SFRD($z$) and $A_{\rm FUV}(z)$ arising from the assumed metallicity and IMF. 

In addition to this, we have also included some of the diffuse emission from the IGM clouds.
We model the He~{\sc ii} Ly-$\alpha$ and He~{\sc ii} Balmer continuum recombination emission 
following the prescription given in HM12 and the Lyman continuum emission due to recombination 
of H~{\sc i} and He~{\sc ii} using the approximations given in \citet{FG09}. 
We do not include the contributions to UVB from He~{\sc i} recombinations and the two photon continuum. 
These contributions are negligible, and if included, can increase  
$\Gamma_{\rm HI}$ by a maximum of 10\% \citep{FG09}. We also do not include the resonance absorption 
of He{\sc ii} which has a negligible effect on $\Gamma_{\rm HI}$, especially at low-$z$ (see HM12).

%
\begin{figure*}
\centering
    \includegraphics[bb=70 370 550 710, width=10cm,keepaspectratio,clip=true]{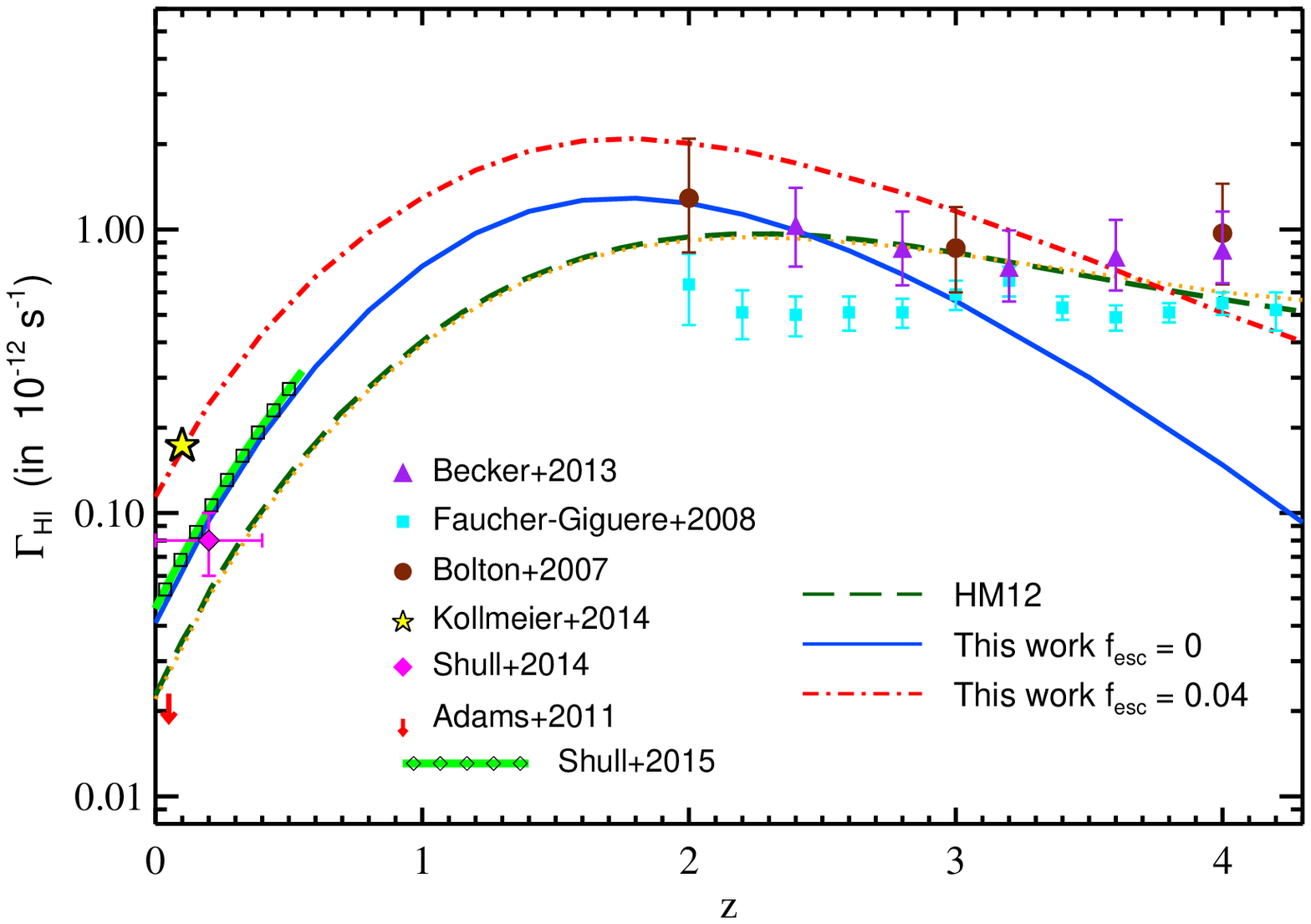}
\caption{The $\Gamma_{\rm HI}$ vs $z$ obtained for our UVB with $f_{\rm esc}=0$ (\emph{solid curve}) 
and with $f_{\rm esc}=4\%$ (\emph{dot-dash curve}) along with the $\Gamma_{\rm HI}$ from 
HM12 (\emph{dash curve}) is plotted. The \emph{dotted curve} shows the $\Gamma_{\rm HI}$ when 
we obtain the UVB using our code with the $\epsilon_{912}^Q(z)$, SED and $f_{\rm esc}$ taken from HM12. 
The $\Gamma_{\rm HI}$ measurements at high-$z$ by \citet{FG08}
(\emph{squares}), by \citet{Bolton07} (\emph{circles}) and by \citet{Becker13}
(\emph{triangles}) are shown. At low-$z$, the lower limit $\Gamma_{\rm HI}$ by \citet{Adams11} 
using non-detection of H$\alpha$ from UGC 7321 \citep[\emph{arrow}; for more details on 
the validity of it see sec. 3.2 of][]{Shull14}, the $\Gamma_{\rm HI}$ which is 
found consistent with the cosmic metal abundances by \citet{Shull14} (\emph{diamond}) and the
inferred $\Gamma_{\rm HI}$ of \citet{Kollmeier14} (\emph{star}) and \citet{Shull15} 
(\emph{green curve with diamonds}) are also plotted. }
\label{fig.3}
\end{figure*}
%

\section{Results and Discussions}\label{sec.res}
Here, we focus on the H~{\sc i} photoionization rate, $\Gamma_{\rm HI}$, obtained using our UVB model. 
This is defined as 
%
\begin{equation}\label{Gama}
\Gamma_{\rm HI}=\int_{\nu_{\rm HI}}^{\infty}d\nu\,\frac{4\pi J_{\nu}}{h\nu}\,\sigma_{\rm HI}(\nu)\,,
\end{equation}
%
where $\nu_{\rm HI}$ corresponds to $\lambda=912$\AA. In Fig.~\ref{fig.3}, we summarize various
available $\Gamma_{\rm HI}$ measurements as a function of $z$. In particular, denoting 
$\Gamma_{\rm HI, 13}=\Gamma_{\rm HI}\times 10^{13}~s^{-1}$, the points of interest
for the present study are $\Gamma_{\rm HI, 13}\sim 1.8$  at $z=0.1$ as inferred 
by \citet{Kollmeier14} which led to the claim of a `photon underproduction crisis' 
and the very recent estimates of $\Gamma_{\rm HI, 13}(z)=0.46(1+z)^{4.4}$ at $z<0.5$ found by \citet{Shull15}. 

To begin with, we validate our code by reproducing the results of HM12. In Fig~\ref{fig.3}, we plot the
$\Gamma_{\rm HI}(z)$ determined by HM12 (\emph{long dashed curve}) and the result of our code obtained 
using the $\epsilon^Q_{\nu}(z)$, SED and $f_{\rm esc}$ used by HM12 (\emph{dotted curve}). Both match with each other 
within $\sim 5$\% accuracy. The minor differences noticed can be attributed to
the different metallicities used and contributions of some of the diffuse emission processes 
ignored in our model. Having validated our code, we use the updated QSO emissivity $\epsilon^Q_{912}(z)$ 
(see Eq.~\ref{Eq.Eqso}) and the $\epsilon^G_{\nu}(z)$ mentioned above (in Section 3.2) to calculate the UVB (and hence
$\Gamma_{\rm HI}$) for different values of $f_{\rm esc}$.

When we use only the QSOs as the source of the UVB (by taking $f_{\rm esc}=0$) and use our updated $\epsilon^Q_{\nu}(z)$, 
we get the $\Gamma_{\rm HI}$ at $z<0.5$ very close (i.e within 10\%) to the values predicted by \citet{Shull15} and 
\citet{Shull14}. We find that the one-sided ionizing flux $\Phi_0$, as defined in \citet{Shull15} is to be $5030$ cm$^{-2}$ s$^{-1}$ 
for our UVB at $z=0$ as compared to 5700 cm$^{-2}$ s$^{-1}$ obtained by \citet{Shull15}. However, because of
the statistical uncertainties in the observed $f(N_{\rm HI})$, the $\Gamma_{\rm HI}$ and $\Phi_0$ predicted by \citet{Shull15} 
can be even higher.
Our $\Gamma_{\rm HI, 13}$ values are 0.41, 0.94, 1.9 and 3.3 at $z=0$, 0.2, 0.4 and 0.6, respectively. 
These are $\sim$2 times higher than the corresponding $\Gamma_{\rm HI}^{\rm HM}$ values.
Now, instead of using our $\epsilon^Q_{\nu}(z)$ fitting form, if we take the best fit PLE models given in 
C09 and PD13 (see Fig~\ref{fig.1}) for $z<2.2$, and estimate the UVB by assuming 
$\epsilon^Q_{912}(z)=0$ at $z>2.2$, we get the $\Gamma_{\rm HI, 13}$ at $z=0$ to be 0.48 and 0.39, respectively. 
It shows that, irrespective of our QLF fits and the fitting form, the updated QSO emissivity will
lead to $\Gamma_{\rm HI}\sim 1.7$ to $2.1 \times \Gamma_{\rm HI}^{\rm HM}$.
Therefore, we conclude that the $\Gamma_{\rm HI}$ inferred by \citet{Shull14} and \citet{Shull15} 
can be explained by the QSOs alone without requiring 
any significant contribution from the galaxies (i.e with $f_{\rm esc}=0$). This is consistent with 
many low-$z$ upper limits on average $f_{\rm esc}$ measured in samples of galaxies   
\citep{Siana07, Cowie09, Siana10, Bridge10, Leitet13}. {\it Therefore, there is no real
photon underproduction crisis when we consider the $\Gamma_{\rm HI}$ measurements
of \citet{Shull15}}.

In our UVB calculations with $f_{\rm esc}=0$, we use a different QSO SED and an updated 
$\epsilon^Q_{\nu}(z)$ as compared to HM12. However, since the $\Gamma_{\rm HI}\propto(3+\alpha)^{-1}$, changing 
$\alpha$ from 1.57 (HM12) to 1.4 at $\lambda<912$\AA~increases the 
$\Gamma_{\rm HI}$ by only 4\%. The main difference in $\Gamma_{\rm HI}$ between 
our UVB and that of HM12 arises because of the updated $\epsilon^Q_{\nu}(z)$. 
It is important to realize that even though the $\epsilon^Q_{912}$ used by us matches with $\epsilon^Q_{912}$
of HM12 at $z=0$, the local UVB is contributed more by ionizing photons coming from 
high-$z$, up to $z\sim2$, where the mean free path for H~{\sc i} ionizing photons is very large
and $\epsilon^Q_{912}(z)$ peaks.

Next we explore the $f_{\rm esc}$ requirements in order to reproduce the 
$\Gamma_{\rm HI}$ inferred by \citet{Kollmeier14}. For simplicity we run
models keeping $f_{\rm esc}$ constant over the full $z$ range. We find  
$f_{\rm esc}=4\%$ is needed to get the  
$\Gamma_{\rm HI, 13}=1.8$ at $z=0.1$ (see Fig~\ref{fig.3}).
Interestingly, the $f_{\rm esc}$ needed in our calculations is much less than
the $f_{\rm esc}=15\%$ required in the HM12 UVB model. Apart from 
2 times higher QSO emissivity, it is partly because of our $\sim$ 3 times higher 
low-$z$ SFRD as compared to HM12. Note that, the value of $f_{\rm esc}\sim 0.02$\% used by
HM12 to estimate UVB at $z=0.1$ is extremely small compared to various low-$z$ observations. 
In passing, we note that for our models with different combinations of 
SFRD and $A_{\rm FUV}$ explored for different extinction curves in 
KS14, we require $f_{\rm esc}$ values similar to or less than what we have 
obtained here for our fiducial model.

In order to compare with observations, we use a relative escape fraction, $f_{\rm esc, rel}$, defined as 
$f_{\rm esc, rel}=f_{\rm esc}\times 10^{0.4A_{\rm FUV}}$. To match the $\Gamma_{\rm HI}$ of K14, 
the model of HM12 that assumes $A_{\rm FUV}=1$ at $z<2$, will require $f_{\rm esc, rel}=38\%$
while we need only $f_{\rm esc, rel} = 15\%$ for our fiducial LMC2 model at $z=0$ (where we determined
$A_{\rm FUV}=1.42$). This $f_{\rm esc, rel}=15\%$ is about a factor $\sim2$ higher than the 
low~$z$ upper limits on $f_{\rm esc, rel}$ given in various studies of galaxy samples as mentioned above. 
However the $f_{\rm esc}$ observed in individual galaxies \citep{Borthakur14} 
and many theoretical estimates \citep[e.g.][]{Kimm14, Roy14} are consistent with it. {\it Therefore,
we conclude that with the updated QSO and galaxy emissivities presented here, even if we wish to generate 
$\Gamma_{\rm HI}$ inferred by K14, the required $f_{esc}$ of ionizing photons from star forming galaxies is
not abnormally high enough to warrant an alternate non-standard source of the UVB.} 

Interestingly, our updated QSO emissivity alone can reproduce the $\Gamma_{\rm HI}$ measurements 
at high $z$ \citep{Bolton07, Becker13} up to $z\sim 2.7$. However, $f_{\rm esc}=4\%$ gives 
a $\Gamma_{\rm HI}(z)$ which marginally overestimates the $\Gamma_{\rm HI}$ measurements at $2<z<3$
(see Fig.3). Irrespective of the low-$z$ $\Gamma_{\rm HI}$, at $z>3$ one needs galaxies to 
contribute more to the UVB \citep[however see,][]{Giallongo15}. 
At high-$z$, using the observations of H~{\sc i} and He~{\sc ii} Ly-$\alpha$ forest,  
it will be possible to constraints the $f_{\rm esc}$ from galaxies \citep[see,][]{Khaire13}.
We plan to do this in the near future.

{\begin{table*}
\begin{center}
\begin{minipage}{160mm}
\caption{Details of observed QLF used to get $\epsilon^Q_{912}$ in our study. }
\begin{tabular}{l c c c c c c c c c }
\hline              
Reference & $\lambda_{\rm band}$&z     &log$\phi^*$& M$^*$& $\gamma_1$  & $\gamma_2$  & log$\epsilon^Q_{\lambda_{\rm band}}$ &   log$\epsilon^Q_{912}$& $\alpha'$ \\
(1) & (2) & (3) & (4) & (5) & (6) & (7) & (8) & (9) & (10)\\
\hline                                                                                                                                               
\citet{Schulze09} & 4450\AA         & 0.15 & -4.81 & -19.46 & -2.0                    & -2.82               & 24.30    &23.83$\pm 0.20$       & 0.5$^{\star}$\\
\citet{Croom09}   & 4686\AA         & 0.54 & -5.98 & -24.10 & -1.4$\pm 0.1$           & -3.5$\pm 0.1$       & 24.59 &24.14$\pm 0.05$           & 0.3 \\
PD13$^{^{\dagger}}$&  4686\AA        & 0.87 & -5.74 & -24.68 & -1.4$\pm 0.1$           & -3.5$\pm 0.1$       & 25.07 &24.59$\pm 0.05$           & 0.5 \\
                  &                 & 1.25 & -5.81 & -25.65 &                         &                     & 25.38 &24.90$\pm 0.05$           &     \\
                  &                 & 1.63 & -5.80 & -25.80 &                         &                     & 25.45 &24.98$\pm 0.05$           &     \\
                  &                 & 2.01 & -6.00 & -26.71 &                         &                     & 25.62 &25.17$\pm 0.05$           &     \\           
\citet{Ross13}    &  7480\AA         & 2.25 & -5.83 & -26.45 & -1.3$^{+0.5}_{-0.2}$    & -3.5$^{+0.3}_{-0.2}$& 25.64 &25.06$^{+0.11}_{-0.16}$   & 0.5 \\
                  &                  & 2.35 & -5.98 & -26.55 &                         &                     & 25.53 &24.95$^{+0.11}_{-0.16}$   &      \\
                  &                  & 2.45 & -6.12 & -26.81 &                         &                     & 25.50 &24.91$^{+0.11}_{-0.16}$   &     \\
                  &                  & 2.55 & -6.12 & -26.77 &                         &                     & 25.47 &24.89$^{+0.11}_{-0.16}$   &     \\
                  &                  & 2.65 & -6.21 & -27.06 &                         &                     & 25.51 &24.92$^{+0.11}_{-0.16}$   &      \\
                  &                  & 2.75 & -6.26 & -27.27 &                         &                     & 25.54 &24.96$^{+0.11}_{-0.16}$   &      \\
                  &                  & 2.90 & -6.29 & -27.22 &                         &                     & 25.49 &24.91$^{+0.11}_{-0.16}$   &     \\
                  &                  & 3.12 & -6.38 & -27.13 &                         &                     & 25.36 &24.79$^{+0.11}_{-0.16}$   &      \\
                  &                  & 3.37 & -6.66 & -27.63 &                         &                     & 25.29 &24.72$^{+0.11}_{-0.16}$   &      \\                            
\citet{Glikman11}  &1450\AA & 4.0  & -5.89 & -24.10 & -1.6$^{+0.8}_{-0.6}$    & -3.3$\pm 0.2$      & 24.80   &24.62$^{+0.50}_{-0.30}$   & NA$^{\ddagger}$\\ 
\citet{Masters12}  &1450\AA & 4.25 & -7.12 & -25.64 & -1.72$\pm 0.28$         & -2.6$\pm 0.63$      & 24.32   &24.13$^{+0.50}_{-0.22}$   &  NA   \\ 
\citet{McGreer13}  &1450\AA & 5.0  & -8.47 & -27.21 & -2.03$^{+0.15}_{-0.14}$ & -4.0                & 23.78   &23.60$^{+0.15}_{-0.14}$   &  NA   \\ 
\citet{Kashikawa15}&1450\AA & 6.0  & -8.92 & -26.91 & -1.92$^{+0.24}_{-0.19}$ & -2.81               & 23.13   &22.94$\pm 0.15$           &  NA   \\
                                                           
\hline                                                            
\end{tabular}                                                     
\hfill
\label{lf_data}
\end{minipage}
\end{center}
\begin{flushleft}
\footnotesize {Column (4) gives $\phi^*$ in units Mpc$^{-3}$ mag$^{-1}$, column (5) gives M$^*$ in 
AB magnitudes and column (8) and (9) gives $\epsilon^Q_{\nu}$ in units erg s$^{-1}$ Hz$^{-1}$ Mpc$^{-3}$.
The $\lambda_{\rm band}$ at 1450\AA, 4450\AA, 4686\AA~and 7480\AA~ corresponds to FUV, B, $g$ and $i$ band, respectively.}
\footnotesize {$^{\star}$We assume $\alpha'=0.5$ consistent with the $k$-correction of \citet{Schulze09}.}
\footnotesize {$^{\dagger}$PD13 stands for \citet{Palanque13}.}
\footnotesize {$^{\ddagger}$NA indicates that the $K$-correction is not applied.}\\
\end{flushleft}
\label{table.main}
\end{table*}
%
\section{Summary}\label{sec.sum}
The recent claim of a `photon underproduction crisis' \citep{Kollmeier14} requires the low-$z$ 
$\Gamma_{\rm HI}$ to be 5 times higher than the one obtained by the UVB model of HM12. 
A similar investigation performed by \citet{Shull15} finds a lower $\Gamma_{\rm HI}$ which is still
2 times higher than that of HM12. Here, we present an
updated  H~{\sc i} ionizing QSO emissivity by using recent QLF measurements.
It turns out that this emissivity is a factor of 1.5 to 2 times higher than what is used by 
HM12 at $0.5<z<2.5$. We estimate the UVB using this emissivity with the help of a radiative transfer 
code developed by us. We show that QSOs alone can give  a factor 2 
required by \citet{Shull15}. Using our updated SFRD which is $\sim$3 times higher than HM12 at low-$z$, 
to get the $\Gamma_{\rm HI}$ predicted by \citet{Kollmeier14} we
require only 4\% of the ionizing photons generated by galaxies to escape into the IGM. Therefore,
there is no need to look for additional sources of ionizing photons 
such as hidden QSOs or decaying dark matter particles. 

\section*{acknowledgments} 
\footnotesize{
We thank M. Shull, D. Weinberg, F. Haardt, A. Paranjape, T. R. Choudhury and H. Padmanabhan for useful comments on the 
manuscript. VK acknowledges support from CSIR.}
\def\apj{ApJ}%
\def\mnras{MNRAS}%
\def\aap{A\&A}%
\def\apjl{ApJ}
\def\aj{AJ}
\def\physrep{PhR}
\def\apjs{ApJS}
\def\pasa{PASA}
\def\pasj{PASJ}
\def\pasp{PASP}
\def\nat{Natur}
\def\araa{AR\&A}
\def\aplett{Astrophysical Letters}

\footnotesize{
  \bibliographystyle{mn2e}
  \bibliography{vikrambib}
}
\end{document}